\documentstyle[prl,aps,preprint]{revtex}

\tighten

\begin{document}
\draft
 
\pagestyle{empty}

\preprint{
\noindent
\begin{minipage}[t]{3in}
\begin{flushleft}
March 1996
\end{flushleft}
\end{minipage}
\hfill
\begin{minipage}[t]{3in}
\begin{flushright}
LBL--38534 \\
UCB--PTH--96/09 \\
\end{flushright}
\end{minipage}
}

\title{Probing Lepton Flavor Violation at Future Colliders
\footnote{This work was supported in part by the Director, Office of
Energy Research, Office of High Energy and Nuclear Physics, Division of
High Energy Physics of the U.S. Department of Energy under Contract
DE-AC03-76SF00098 and in part by the National Science Foundation under
grant PHY-95-14797.}}

\author{Nima Arkani-Hamed, 
Hsin-Chia Cheng,
Jonathan L. Feng
\thanks{Research Fellow, Miller Institute for Basic Research in
Science.},
and
Lawrence J.~Hall
}

\address{Theoretical Physics Group,\\
        Lawrence Berkeley National Laboratory\\ 
        and\\ Department of Physics\\
       University of California, Berkeley, California 94720
}

\maketitle

\begin{abstract}
Supersymmetric theories with significant lepton flavor violation have
$\tilde{e}$ and $\tilde{\mu}$ nearly degenerate. In this case,
pair production of $\tilde{e}^+ \tilde{e}^-$ and
$\tilde{\mu}^+ \tilde{\mu}^-$ at LEPII and at the 
Next Linear Collider
leads to the phenomenon of slepton oscillations, which is 
analogous to neutrino
oscillations. The reach in $\Delta
m^2$ and $\sin^2 2 \theta$ gives a probe of lepton flavor 
violation which is
significantly more powerful than the current bounds from rare processes,
such as $\mu \to e\gamma$.  Polarizable $e^-$ beams and the
$e^-e^-$ mode at the NLC are found to be promising options.
\end{abstract}



Two fundamental questions of particle physics are the origin of
electroweak symmetry breaking and the pattern of the quark and lepton
mass matrices, known as the flavor problem.  Extensions of the standard
model with weak scale supersymmetry have been widely studied: the
enhanced spacetime symmetry offers an understanding of the hierarchy of
the weak and Planck scales, electroweak symmetry breaking is triggered
by the dynamics of a heavy top quark, and the unification prediction for
the weak mixing angle is highly successful if there are superpartners at
the weak scale. The experimental discovery of superpartners would
represent enormous progress in understanding electroweak symmetry
breaking, but would it allow progress on the flavor problem?

It is probable that all the quarks and leptons have been
discovered. Although further light will be shed on the flavor problem by
measuring the 13 fermion masses and mixings to greater accuracy, within
the minimal three generation standard model there are no new flavor
parameters to measure. In any supersymmetric extension of the standard
model, the superpartners of the quarks and leptons must be given
masses. There are 15 new flavor parameters in the scalar mass
eigenvalues, and seven new flavor mixing matrices\cite{ACH}

\begin{equation}
W_a = U_a^\dagger V_a \ , \quad a = u_{L,R}, d_{L,R}, e_{L,R}, \nu_L \ ,
\end{equation}
arising as relative rotations between the matrices $U_a$ and $V_a$ that
diagonalize the scalar and fermion mass matrices, respectively. At the
neutral gaugino vertex of species $a$, the $i$-th generation scalar is
converted to the $j$-th generation fermion with amplitude
${W_a}_{ij}$. If supersymmetry is correct, it will furnish a large new
arena for studying the problem of flavor.

The last twenty years have seen a successsion of discoveries of heavy
quarks and leptons. After the initial discovery and mass measurement,
the focus has turned to measurement of the flavor violations via the
parameters of the Cabibbo-Kobayashi-Maskawa matrix $V_{CKM} =
V_{u_L}^\dagger V_{d_L}$.  If supersymmetry is discovered, we envisage a
similar progression: after measurements of superpartner masses, the
focus will shift to a study of the new flavor mixing matrices at the
gaugino vertices.  Rare flavor changing processes, such as $\mu
\to e \gamma$, $\tau \to \mu \gamma$, $\tau \to e \gamma$, 
$b \to s \gamma$, and neutral meson mixing, provide important
constraints on the $W_a$ mixing matrices via the virtual effects of
superpartners. In this paper, however, we show that, once superpartners
are discovered, it will be possible to probe these matrices much more
powerfully by directly observing the change in flavor occurring at the
superpartner production and decay vertices.  We consider lepton flavor
violation, and find that, if sleptons are made at LEP II or the NLC, the
21, 32, and 31 elements of $W_{e_L,e_R}$ will be probed considerably
beyond the most stringent present limits, which result from $\mu \to e
\gamma$, $\tau \to \mu \gamma$, and $\tau \to e \gamma$, respectively.

There are many theoretical ideas for the origin of the scalar and
fermion masses in supersymmetry and the symmetries that govern
them\cite{TOF}.  In this letter we do not discuss these theories; we
concentrate on the question of how well $W_{e_L,e_R}$ can be probed at
future electron colliders in a model independent way, assuming only that
sleptons are directly produced. Nevertheless, an important question is
which experimental signature will provide the best probe of this
physics, and hence is most likely to produce a signal. To evaluate this,
we find the reach of NLC for ${W_{e_L,e_R}}_{ij}$, $i \neq j$, and
compare it to the corresponding CKM matrix element, ${V_{CKM}}_{ij}$. We
find that only in the case $ij = 12$ can the NLC probe mixing angles as
small as the CKM case, and in this case the probe can be very far
beneath the CKM case. Hence in this paper we limit ourselves to an
analysis of the lightest two generations. Further details of this case,
and the reach for flavor violation involving the tau, will be given in a
subsequent paper\cite{ACFH}.

If ${W_{e_L,e_R}}_{12}$ are comparable to the Cabibbo angle, then the
rate for $\mu \to e \gamma$ is typically four orders of magnitude above
the experimental bound; this is part of the well-known supersymmetric
flavor changing problem\cite{HKT}. It is solved by having considerable
degeneracy between the superpartners $\tilde{e}$ and $ \tilde{\mu}$,
leading to a superGIM cancellation in the amplitude for $\mu \to e
\gamma$. The near degeneracy of $\tilde{e}$ and $ \tilde{\mu}$, together
with their mass mixing, which results in non-zero ${W_{e_L,e_R}}_{12}$,
implies that the direct production of $\tilde{e}$ and $ \tilde{\mu}$
results in lepton flavor oscillations, analogous to strangeness
oscillations and neutrino oscillations.  Unlike the neutrino case,
however, $\tilde{e}$ and $ \tilde{\mu}$ decay very quickly, and hence
the relevant signal is the time integrated one.  Nevertheless, the reach
of an experiment is best described by plotting event rate contours in
the $(\sin 2 \theta, \Delta m^2)$ plane\cite{KRAS}.

The gauge eigenstate scalars $|{\tilde{e}}\rangle$, 
$|{\tilde{\mu}}\rangle$ are
related to the mass eigenstate scalars $|{1}\rangle, |{2}\rangle$ via

\begin{eqnarray}
|{\tilde{e}}\rangle &=& + \cos \theta |{1}\rangle + \sin \theta |{2}\rangle
\nonumber \\
|{\tilde{\mu}}\rangle &=& - \sin \theta |{1}\rangle + 
\cos \theta |{2}\rangle \ , 
\end{eqnarray}
where $\sin \theta = W_{12}$.  Suppose that at time $t=0$ we produce a
gauge eigenstate selectron in an $e$--$e$ collision: $|{\psi(0)}\rangle =
|{\tilde{e}}\rangle$. The state at time $t$ is

\begin{eqnarray}
|{\psi (t)}\rangle &=& \cos \theta 
e^{-{\Gamma\over 2} t - i m_{1} t} |{1}\rangle + 
\sin \theta e^{-{\Gamma\over 2} t - i m_{2} t} |{2}\rangle \nonumber \\
&=& (\cos^{2} \theta e^{-{\Gamma \over 2} t - i m_{1} t} + \sin^{2} \theta
e^{-{\Gamma \over 2} t - i m_{2} t}) |{\tilde{e}}\rangle \nonumber \\
&& - \sin \theta 
\cos \theta (e^{-{\Gamma \over 2} t - i m_{1} t} -
e^{-{\Gamma \over 2} t - i m_{2}t}) |{\tilde{\mu}}\rangle \ ,
\end{eqnarray}
where we have neglected the difference between the widths of the two
mass eigenstates and set them equal to $\Gamma$.  The probability
$P(\tilde{e} \to f_{\mu})$ that the gauge eigenstate selectron
decays into the final state containing a muon, $f_{\mu}$, is

\begin{eqnarray}
P(\tilde{e} \to f_{\mu}) = {\int_{0}^{\infty} dt 
|\langle{\tilde{\mu}}|{\psi (t)}\rangle|^{2} \over
\int_{0}^{\infty} dt 
\langle{\psi(t)}|{\psi(t)}\rangle} \times B(\tilde{\mu} \to
f_{\mu}) \nonumber \\
= 2 \sin^{2} \theta \cos^{2} \theta {(\Delta m^{2})^{2} 
\over 4 \bar{m}^{2} \Gamma^{2} + (\Delta m^{2})^{2}} \times
B(\tilde{\mu} \to
f_{\mu}) \ ,
\end{eqnarray}
where $\Delta m^{2} = m_{1}^{2} - m_{2}^{2}$, $\bar{m} =
(m_{1} + m_{2})/2$, and $B(\tilde{\mu} \to f_{\mu})$ is the
branching fraction for $\tilde{\mu} \to f_{\mu}$. The term depending
on $\Delta m^{2}$ is the quantum interference factor neglected in\cite
{KRAS}. Note that when $\Delta m^{2} \gg \bar{m}
\Gamma$, this factor becomes 1 and the interference effect can be
ignored.  However, when $\Delta m^{2} \ll \bar{m} \Gamma$, the factor
goes to zero and the interference effect cannot be neglected.  $\Gamma$
is typically much smaller than $\bar{m}$: for instance, if the only
decay mode for a gauge eigenstate right-handed selectron is $\tilde{e}
\to e \tilde{\chi}^0$, 
where $\tilde{\chi}^0$ the lightest neutralino, $\Gamma / \bar{m} = 
{\alpha \over 2 \cos^{2} \theta_{W}} \left( 1 - m^{2}_{\tilde{\chi}^0} /
\bar{m}^{2}\right)^{2} \sim 0.01$. Thus, as long as $\Delta m^{2}/ 
\bar{m}^{2} > 0.01$, there is no interference suppression of the 
flavor changing process. However, $B(\mu \to e \gamma)$ constrains the
product $\sin \theta \cos \theta \Delta m^{2}/\bar{m}^{2}$ to be
(roughly) less than $ \sim 0.01$, so there is competition between these
different probes of flavor violation.

We have calculated the cross sections for the flavor-violating processes
$e^+e^- \to e^{\pm} \mu^{\mp} \tilde{\chi}^0 \tilde{\chi}^0$ 
and $e^-e^- \to e^- \mu^- 
\tilde{\chi}^0 \tilde{\chi}^0$. In calculating these cross sections, we 
have correctly treated the different interferences in different channels.
In the
$e^-e^-$ case, the amplitude comes from $t$-channel neutralino exchange
producing gauge eigenstate selectrons, while in the $e^+e^-$ case there
are additional contributions from $s$-channel annihilation into
$\gamma/Z$ producing gauge eigenstate selectrons and smuons. In both
cases the sleptons produced oscillate and decay into leptons and lighter
superpartners.  Consider the $e^-e^-$ case with both beams right
polarized.  The cross section for $e_{R}^{-} e_{R}^{-} \to e^{-}
\mu^{-} \tilde{\chi}^0\tilde{\chi}^0$ 
depends on the right-handed mixing angle in the 
combination $\sin 2\theta_{R}$, on the mass difference of the
right-handed scalars $\Delta m^{2}_{R}/\bar{m}^{2}_{R}$ (via the
interference effect), on the average mass of the right-handed selectrons
and smuons $\bar{m}^{2}_{R}$, and finally, assuming that the lightest
superpartner (LSP) is pure Bino, on the Bino mass $M_{1}$. Fixing
$\bar{m}$ and $M_{1}$, we will give contour plots below for the cross
section in the $(\sin 2 \theta_R, \Delta m^{2}_{R}/\bar{m}^{2}_{R})$
plane. For comparison, we will also include contours of $B(\mu \to e
\gamma)$ in our plots. The $\theta_R$-dependent 
amplitude for $\mu \to e \gamma$ contains
two pieces: one depending on the same parameters just discussed and the
other depending further on the left-right scalar mass mixing parameter
$A + \mu \tan \beta$ and the left-handed scalar masses.
       
Having discussed the flavor-violating cross sections, we now examine the
possibility of detecting flavor-violating signals at future colliders.
We first consider the sensitivity of the LEP II $e^+e^-$ collider, with
a center of mass energy $\sqrt{s} = 190 \text{ GeV}$ and an integrated
luminosity of roughly $500 \text{ pb}^{-1}$.  We then turn to the NLC,
with design energy $\sqrt{s} = 500 \text{ GeV}$ and luminosity $50\text{
fb}^{-1}/\text{yr}$ in $e^+e^-$ mode.  The $e^-e^-$ luminosity is
currently being studied\cite{Bauer}, and may be degraded somewhat from
the $e^+e^-$ luminosity.  We will, however, assume an event sample of $50
\text{ fb}^{-1}$ in both modes.  A $5\sigma$ discovery signal then
requires $S \ge 7.1 \sqrt{B} \sqrt{0.5/{Y}}$ for LEP II and $S \ge 0.71
\sqrt{B} \sqrt{50/{Y}}$ for the NLC, where $S$ and $B$ are the signal
and background cross sections after cuts (in fb), and the total
integrated luminosity is $Y \text{ fb}^{-1}$.

To discuss the flavor violation discovery potential of LEP II, we first
choose some representative values for the various SUSY parameters.
(Some implications of deviations from these choices will be discussed
below.)  Sleptons with mass below 85--90 GeV are expected to be
discovered at LEP II. We therefore consider the case where
$m_{\tilde{e}_R} \approx m_{\tilde{\mu}_R} \approx 80 \text{ GeV}$.  The
LSP must be lighter than this, and we assume that it is Bino-like with
mass $M_1 = 50 \text{ GeV}$.  For simplicity, we also assume that the
production of all other supersymmetric particles is suppressed, either
kinematically, or, for example, in the case of neutralinos, through
mixing angles.  The sleptons, then, decay directly to the LSP, and the
flavor-violating signal is $e^+e^- \to (\tilde{e}_R,
\tilde{\mu}_R) (\tilde{e}_R, \tilde{\mu}_R) \to 
e^{\pm} \mu^{\mp} \tilde{\chi}^0
\tilde{\chi}^0$.

\begin{figure}
\psfig{file=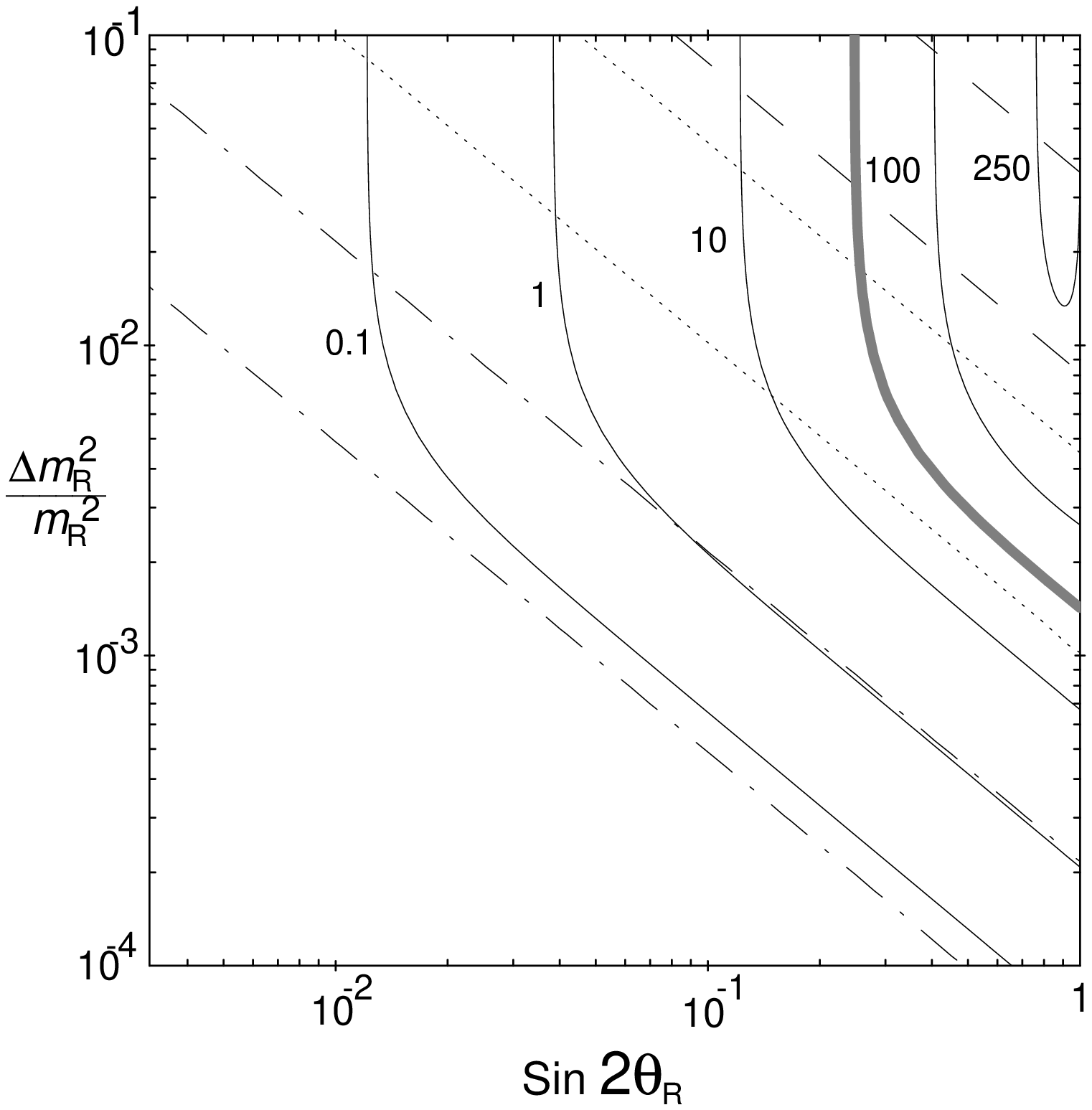,width=0.96\textwidth}
\caption{Contours of constant $\sigma ( e^+e^- \to e^{\pm} \mu^{\mp} 
\tilde{\chi}^0\tilde{\chi}^0 )$ (solid) in fb
for LEP II, with  $\protect\sqrt{s} = 
190 \text{ GeV}$, $m_{\tilde{e}_R}, m_{\tilde{\mu}_R} \approx 80
\text{ GeV}$, and $M_1 = 50 \text{ GeV}$.
The thick gray contour represents the optimal experimental reach
in one year (40 fb).
Constant contours of $B(\mu \to e\gamma)=4.9\times 10^{-11}$ and 
$2.5\times 10^{-12}$ are also plotted for
degenerate left-handed sleptons with mass 120 GeV and $\tilde{t} \equiv
-(A + \mu \tan\beta)/\bar{m}_R = 0$ (dotted), 2 (dashed), and 50
(dot-dashed).
\label{fig:LEPII}}
\end{figure}

At LEP II energies, the dominant standard model background to the
$e^{\pm} \mu^{\mp}$ final state is $W$ pair production, where both $W$
bosons decay to $e$ or $\mu$, either directly or through $\tau$ leptons.
Including branching ratios, this cross section is 680 fb.  The $WW$
background may be reduced with cuts, as has been discussed in a number
of studies\cite{LEPII}. (Of course, if sleptons are significantly
lighter than 80 GeV, one may run below $\sqrt{s} = 160 \text{ GeV}$ and
eliminate $WW$ production altogether.) Depending on the LSP mass, the
cuts may be optimized to reduce the background to $\sim$10--100 fb,
while retaining 40\% -- 60\% of the signal.  Given an integrated
luminosity of $500 \text{ pb}^{-1}$, then, the required cross section
for a $5\sigma$ effect is $\sim$40--185 fb.  The flavor-violating cross
section is plotted in Fig.~\ref{fig:LEPII}, along with the constraint
from $B(\mu\to e\gamma)$ for various values of $\tilde{t}\equiv
-(A+\mu\tan\beta)/\bar{m}_R$.  (In the limit of large left-handed scalar
mass, the $A+\mu\tan\beta$ contribution to $\mu \to e \gamma$ vanishes,
and so this limit corresponds to the $\tilde{t} = 0$ contour.) The cross
section contours extend to $\sin\theta_{R}\sim 0.15$, and for low values
of $\tilde{t}$ extend the reach in parameter space significantly.

We have assumed above that no other supersymmetric particles are
produced.  If this is not the case, there may be supersymmetric
backgrounds.  However, the supersymmetric backgrounds tend to be small
relative to the $WW$ background; in the case of stau pairs, for example,
after branching ratios are included, this background is ${\cal O} (10
\text{ fb})$.  We have also assumed above that we are in the region where
the lighter chargino and neutralinos are gaugino-like.  If they are
Higgsino-like, the slepton decay widths are greatly reduced, and the
$\Delta m^2_R /(\bar{m}_R \Gamma)$ suppression takes effect only for smaller
$\Delta m^2_R$.  Thus, the $e^{\pm} \mu^{\mp}$ signal can probe regions of
even smaller $\Delta m^2_R / \bar{m}^{2}_R$. However, for $e^+ e^-$
machines, the cross section become smaller when $M_1$ is large. The reach
in mixing angle may be slightly worse.

If sleptons are not produced at LEP II, they may be discovered at the
NLC.  To consider the potential for discovering slepton flavor violation
there, we consider right-handed slepton masses $m_{\tilde{e}_R}
,\, m_{\tilde{\mu}_R} \approx 200 \text{ GeV}$, and $M_1= 100 \text{
GeV}$.  Again, we assume that we are in the gaugino region, and that the
production of other sparticles is suppressed.

\begin{figure}
\psfig{file=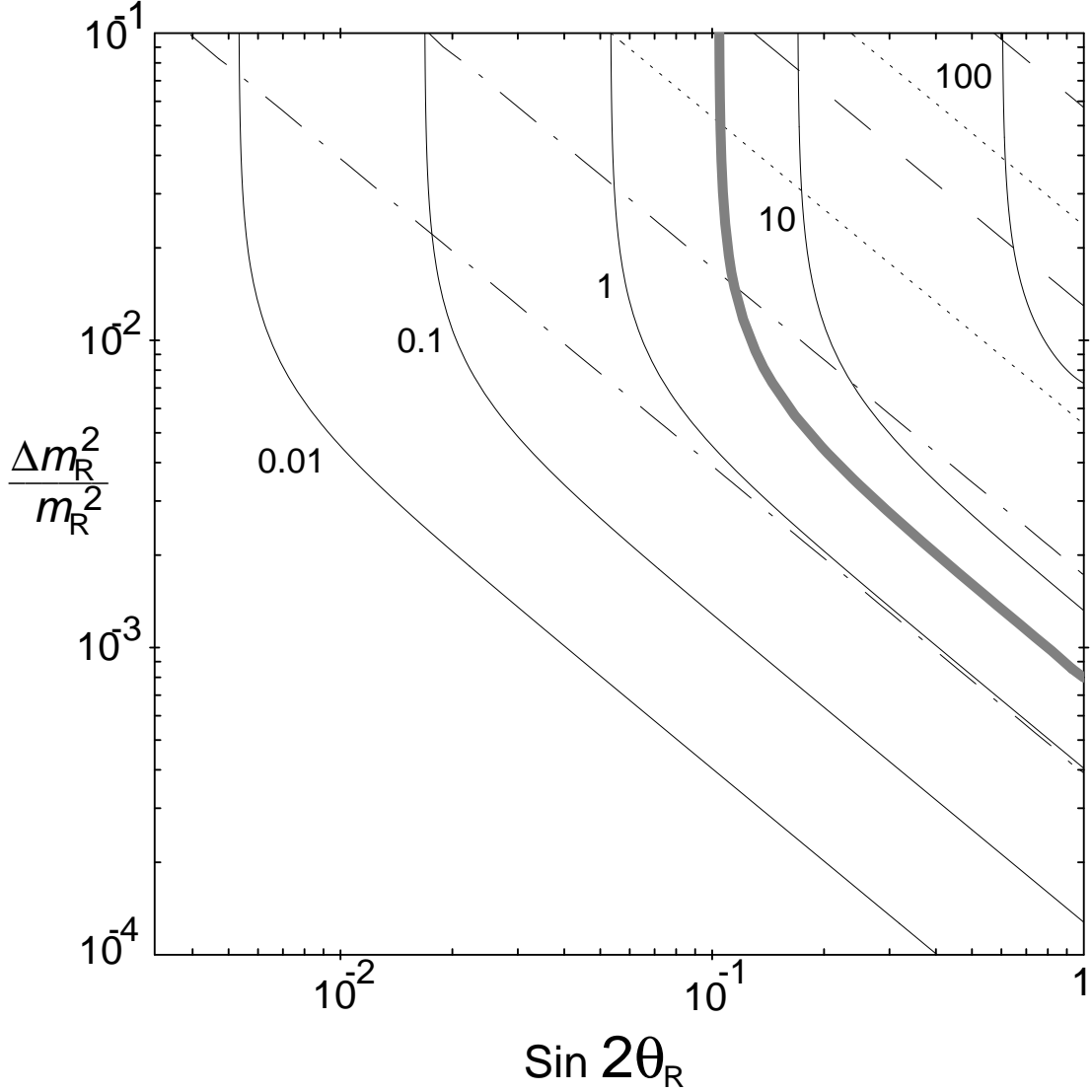,width=0.96\textwidth}
\caption{Contours of constant $\sigma (e^+e^-_R\to e^{\pm} \mu^{\mp} 
\tilde{\chi}^0\tilde{\chi}^0 )$ (solid) in fb for NLC, 
with $\protect\sqrt{s} = 500 \text{ GeV}$, 
$m_{\tilde{e}_R}, m_{\tilde{\mu}_R} \approx 200
\text{ GeV}$, and $M_1 = 100 \text{ GeV}$ (solid).  
The thick gray contour represents the experimental reach in one year.
Constant contours 
of $B(\mu \to e\gamma)$ are also plotted as in
Fig.~\protect\ref{fig:LEPII}, but for left-handed sleptons degenerate
at 350 GeV.
\label{fig:NLC+}}
\end{figure}

There are many options at the NLC, as both highly polarized $e^-$ beams
and $e^+e^-$ and $e^-e^-$ modes may be available.  We consider first the
$e^+e^-$ modes. At NLC energies, $W$ pair production is still a large
background at 7 pb, but there are now others, including $e^{\pm} \nu
W^{\mp}$, which, at 5 pb, is a large background even though the electron
tends to disappear down the beampipe, and $(e^+e^-) W^+ W^-$, which is
only 200 fb, but is difficult to remove from the signal.  Nevertheless,
efficient cuts have been devised for (flavor-conserving) selectron and
smuon pair production\cite{BV,JLC}, and these also effectively isolate
the flavor-violating signal.  Applying the cuts of Ref.~\cite{BV}, we
find that the standard model $e^{\pm}\mu^{\mp}$ background is reduced to
5.2 fb for unpolarized beams, while $\sim$30\% of the signal is
retained.  This may be improved further by using a $e^-_R$ beam, which
doubles the signal and removes $W$ pair production, reducing the
background to 2.6 (2.3) fb for 90\% (95\%) beam polarization. (Note also
that the $e^-_R$ beam also highly suppresses the pair production of
Wino-like charginos.)  Given a year's running at design luminosity, the
required $5\sigma$ signal is 3.8 (3.6) fb.  Cross sections for
$e^+e^-_R \to e^{\pm} \mu^{\mp} \tilde{\chi}^0 \tilde{\chi}^0$ 
at the NLC are given in
Fig.~\ref{fig:NLC+}.  We see that the NLC in $e^+e^-$ mode is also a
powerful probe of the flavor-violating parameter space, extending to
$\sin\theta_{R} = 0.06$ and probing parameter space for which $B(\mu \to
e\gamma) =  10^{-14}$ ($10^{-11}$) for
$\tilde{t} = 2$ (50).  The extent of parameter space
probed is seen to be insensitive to beam polarization.

An intriguing feature of the NLC is its ability to run in $e^-e^-$ mode.
This option allows one to polarize both beams, and has extremely low
backgrounds.  For example, $WW$ production, previously our most
troublesome background, and chargino production are both eliminated.
However, as first noted in Ref.~\cite{KL}, slepton pair production is
allowed, as SUSY theories naturally provide Majorana particles, the
neutralinos, which violate fermion number.  The flavor-violating signal
is slepton pair production followed by decays to the final state
$e^-\mu^- \tilde{\chi}^0 \tilde{\chi}^0$.  
In fact, the backgrounds are so small in $e^-e^-$
mode that $\mu^-\mu^-$ final states may also be used\cite{ACFH}.

\begin{figure}
\psfig{file=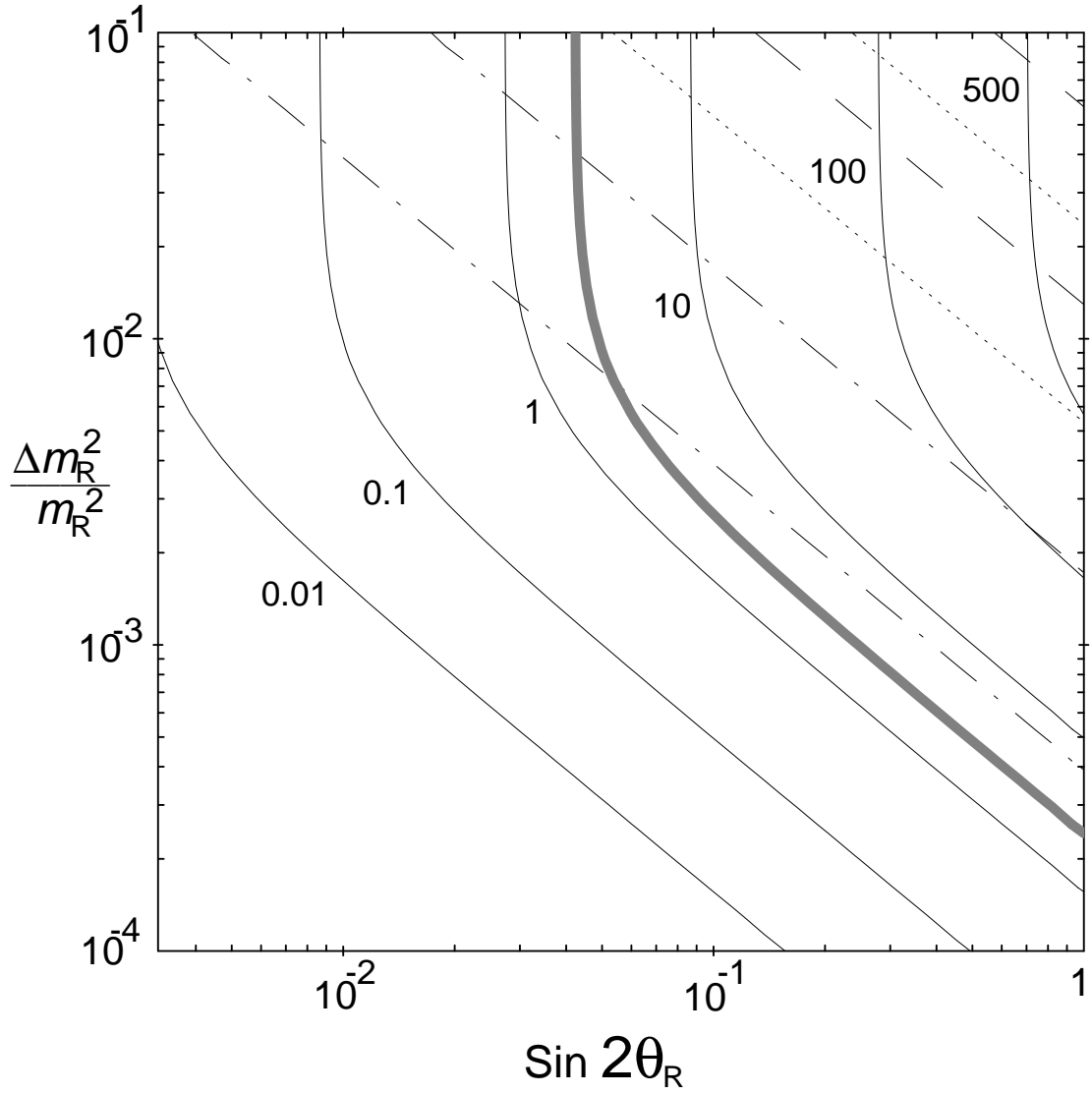,width=0.96\textwidth}
\caption{Same as in Fig.~\protect\ref{fig:NLC+}, but for
$\sigma (e^-_R e^-_R \to e^- \mu^- \tilde{\chi}^0\tilde{\chi}^0 )$.
\label{fig:NLC-}}

\end{figure}

Assuming excellent lepton charge identification and a hermetic detector,
there are essentially no backgrounds for the RR beam polarization.  For
LR (LL), the dominant background is $e^-\nu W^-$ with cross section 43
(400) fb\cite{CC}, where the electron is required to have rapidity $\eta
< 3$ and the branching fraction of $W^-\to \mu^- \bar{\nu}_{\mu}$ has
been included.  Thus, without any additional cuts, if both beams are
90\% (95\%) right-polarized, the background is reduced to 12 (5.1) fb,
and the required $5\sigma$ signal is 2.4 (1.6) fb. Cross sections for
$e^-_R e^-_R \to e^- \mu^- \tilde{\chi}^0 \tilde{\chi}^0$ are given in
Fig.~\ref{fig:NLC-}.  This proves to be the most sensitive mode
considered so far, probing mixing angles with $\sin\theta_{R} = 0.02$
and probing parameter space for which $B(\mu \to e\gamma) = 
10^{-15}$ ($ 10^{-12}$) for $\tilde{t}
= 2$ (50). 

Finally, we note that once lepton flavor violation is detected, the next
step will be to identify its sources and measure it precisely.  For
simplicity, we have chosen scenarios in which the flavor-violating
signal results solely from $W_{e_R}$ mixing.  This analysis may be
applied to $W_{e_L}$ mixing with the analysis of left-handed sleptons.
Of course, in more general settings, flavor-violating signals from both
$W_{e_L}$ and $W_{e_R}$ mixing may be accessible.  For example, if both
left- and right-handed charged sleptons are available, the
flavor-violating cross section in $e^+e^-$ mode may depend on both
mixings.  However, in the $e^-e^-$ mode, one can isolate the flavor
violation to either $W_{e_L}$ or $W_{e_R}$ by polarizing both beams, a
considerable aid in disentangling the flavor-violating matrices.

If sleptons are discovered at the NLC, the $e\mu$ signal will
provide the most powerful probe of flavor violation mixing between the
two lightest generations.  Many theories of flavor will be probed: for
example, those that give $W_{12} = \sqrt{m_e/m_{\mu}}$, in analogy to
$V_{us} = \sqrt{m_d/ m_s}$.  Although it is possible that lepton flavor
is exactly conserved, many unified theories give $W_{12}$ 
and $\Delta m^2$ large enough to
be detected, as, for example, in Ref.\cite{PT}.  
Important tests are also possible
in the third generation: $\tau \to \mu \gamma$ ($\tau \to e \gamma$) do
not give bounds on $W_{32}$ $(W_{31})$, (although $\mu\to e\gamma$ 
constrains their product,) but can be probed down to 
approximately 0.2
(0.05) at the NLC\cite{ACFH}.


We are grateful to H.~Murayama for many useful discussions.
L.J.H. thanks M.~Peskin for asking whether lepton flavor violation
mixing angles might be directly probed at the NLC. This work was
supported in part by the Director, Office of Energy Research, Office of
High Energy and Nuclear Physics, Division of High Energy Physics of the
U.S.  Department of Energy under Contract DE--AC03--76SF00098 and in
part by the NSF under grant PHY--95--14797.  The work of N.~A.-H. is
supported by NSERC.  J.L.F. is supported by a Miller Institute Research
Fellowship.

\input psfig

\noindent

\noindent

\noindent

\end{document}